\documentstyle[12pt]{article}
\newlength{\inda}
\newlength{\indb}

\begin{document}

\markright{Extent of the Immirzi Ambiguity}

\begin{center}
{\large \bf Extent of the Immirzi Ambiguity in Quantum General
Relativity}
\\[3mm] {Guillermo A Mena Marug\'{a}n}
\\[3mm] {\it  Centro de F\'{\i}sica Miguel A. Catal\'{a}n, IMAFF, CSIC,
\\ Serrano 121, 28006 Madrid, Spain}
\end{center}

\begin{abstract}
The Ashtekar-Barbero formulation of general relativity admits a
one-parameter family of canonical transformations that preserves
the expressions of the Gauss and diffeomorphism constraints. The
loop quantization of the connection formalism based on each of
these canonical sets leads to different predictions. This
phenomenon is called the Immirzi ambiguity. It has been recently
argued that this ambiguity could be generalized to the extent of a
spatially dependent function, instead of a parameter. This would
ruin the predictability of loop quantum gravity. We prove that
such expectations are not realized, so that the Immirzi ambiguity
introduces exclusively a freedom in the choice of a real number.
\end{abstract}

Nearly fifteen years ago Ashtekar introduced a description of the
gravitational field in terms of a complex SO(3) connection and a
canonically conjugated densitized triad \cite{Ash,book}. This
connection formalism provides one of the most promising approaches
to quantize general relativity. The use of Ashtekar variables
drastically simplifies the expression of the gravitational
constraints. In addition, the shift from geometrodynamics to
connection dynamics allows the interchange and unification of the
mathematical techniques employed in the quantization of gauge
matter field theories and in quantum gravity.

The main problem of the original formalism proposed by Ashtekar is
that the connection variable is complex. This introduces severe
obstacles for the success of the quantization program. On the one
hand, the real part of the Ashtekar connection coincides with the
SO(3) connection compatible with the densitized triad. In the
quantum theory, this relation between gravitational variables is
encoded in the so-called ``reality conditions'', which are
extremely hard to implement \cite{book,Mat}. On the other hand, no
suitable mathematical tools have been developed to deal with the
complexified SO(3) gauge group, which can be considered
non-compact.

A way to circumvent this problem was suggested by Barbero, who
realized that, by means of a slight modification of the
generalized canonical transformation introduced by Ashtekar on the
gravitational phase space, one reaches in fact a real SO(3)
connection while respecting the canonical symplectic structure
\cite{Fer}. The new canonical variables for gravity are usually
called the Ashtekar-Barbero variables, and have been extensively
employed in the non-perturbative quantization of general
relativity, specially in loop quantum gravity \cite{loop}. The
main drawback of this approach with respect to the original
Ashtekar formulation is that the Hamiltonian constraint loses its
extremely simple form. But this is a minor problem compared with
the operational advantages of dealing with the real SO(3) group
(or, equivalently, the real SU(2) group \cite{book}) as the
relevant gauge group.

The canonical transformation discussed by Barbero can be extended
to a one-parameter family of transformations, all of them
preserving the kinematical structure and leading to a real
connection as the configuration variable \cite{Gio}. By preserving
the kinematical structure we understand that the transformation
does not affect the form of the Poisson brackets nor the
expression of the non-dynamical constraints, namely, the Gauss and
diffeomorphism constraints. The parameter of these canonical
transformations is commonly denoted by $\beta$ and called the
Immirzi parameter. The remarkable point noted by Immirzi is that,
whereas in the classical theory his parameter labels only
different equivalent descriptions of the gravitational phase
space, in the quantum theory there exists an ambiguity, so that
the geometrical predictions depend on the value of $\beta$
\cite{Gio}. This is the case, e.g., of the spectrum of the area
operator \cite{Gio,AL}.

Obviously, the existence of the Immirzi ambiguity implies that the
canonical transformations relating the formulations with different
values of $\beta$ cannot be implemented unitarily in the quantum
theory \cite{RT}. Trying to understand the origins of this
ambiguity, it has been compared with the $\theta$ ambiguity that
arises in Yang-Mill theories \cite{GOP}, although (unlike the
situation found in those theories) it does not appear as a
consequence of a multiply connected configuration space
\cite{Sam}. A key point about the Immirzi ambiguity is that it
affects the physical predictions only to the extent of a constant
parameter. Even in the worst of the theoretical scenarios, namely
if the Immirzi parameter cannot be fixed by any fundamental
principle, a single physical measurement would suffice to
determine which is the actual value of $\beta$ realized in nature.

This is the viewpoint which is more strongly advocated in loop
quantum gravity. In more detail, the standard proposal consists in
removing the Immirzi ambiguity by studying the entropy ${\sc S}$
of a quantum black hole. Ashtekar and his collaborators \cite{QBH}
have shown that ${\sc S}=2\pi\beta_0 {\sc A}/(\beta l^2)$ for
large horizon areas ${\sc A}$, where
$\beta_0=\ln{2}/(\pi\sqrt{3})$ and the canonical Ashtekar-Barbero
variables have an identity Poisson bracket multiplied by $l^2$
\cite{Kras}. Then, if the Bekenstein-Hawking formula ${\sc S}={\sc
A}/(4 l_p^2)$ is verified in nature, we must have $\beta l^2=
8\pi\beta_0 l_p^2$. Here $l_p$ denotes the Planck length in
low-energy physics. Since it is generally assumed that $l^2=8\pi
l_p^2$, one gets $\beta=\beta_0$. Let us comment, nevertheless,
that the length scales $l$ and $\sqrt{8\pi}l_p$ might be
different, as has been remarked by Rainer \cite{Martin}.

It has been recently argued that the extent of the Immirzi
ambiguity may in fact be generalized from a parameter to a
function of the spatial position \cite{Samuel}. If this were
possible, any prediction in loop quantum gravity, including the
expression of the black hole entropy, would contain a free
function of the spatial coordinates. For the particular case of
the black hole entropy, it would be difficult to regain the
Bekenstein-Hawking formula unless $\beta$ is a constant, because
so are all the quantities involved in this formula. But much more
importantly, in order to check the validity of the quantum
predictions and remove the ambiguity, one would have to perform an
infinite number of measurements. This would ruin the
predictability of loop quantum gravity. The aim of the present
work is to prove that such expectations are not fulfilled, and
that the freedom introduced by the Immirzi ambiguity consists only
in a constant parameter.

Let us start by considering vacuum general relativity in the
Ashtekar-Barbero formulation. The gravitational variables can be
chosen as a SO(3) connection $A_a^i$ and a densitized triad
$\tilde{E}^a_i$. Lowercase Latin letters from the beginning and
middle of the alphabet denote spatial and SO(3) indices (running
from 1 to 3), respectively, the latter being lowered and raised
with the identity metric. The inverse metric on the spatial
sections is
$h^{ab}(\tilde{E})=\tilde{E}^a_i\tilde{E}^{bi}/\tilde{E}^2$, with
$\tilde{E}=\sqrt{ {\rm det} \tilde{E}^a_i}$. The connection
$A_a^i$ can be expressed in terms of the SO(3) connection
compatible with the triad $\Gamma_a^i(\tilde{E})$ and the
extrinsic curvature in triadic form $K_a^i$:
\begin{equation} \label{AB}
A_a^i=\Gamma_a^i(\tilde{E})-\beta K_a^i,\end{equation} where
$\beta$ is a constant. Remember that, given a densitized triad and
its inverse $E_{_{_{\!\!\!\!\!\!\sim}}\;a}^i$,
\begin{equation} \label{spin}
\Gamma_a^i(\tilde{E})= \epsilon^{ijk}
\tilde{E}^b_j\left(\partial_{[b}E_{_{_{\!\!\!\!\!\!\sim}}\;a]k}+
E_{_{_{\!\!\!\!\!\!\sim}}\;a[l}\tilde{E}^c_{k]}
\partial_bE_{_{_{\!\!\!\!\!\!\sim}}\;c}^l\right).
\end{equation}
The indices in square brackets are anti-symmetrized. When
$\beta=1$, $A_a^i$ reproduces the real connection introduced by
Barbero \cite{Fer}, that we will distinguish with the notation
$\underline{A}_a^i$. The original, complex connection of the
Ashtekar formalism is obtained with $\beta=i$ \cite{book}. In
general, $A_a^i$ is a real connection for any real number $\beta$
because, in Lorentzian gravity, one can always choose $\Gamma_a^i$
to be a real SO(3) connection and $K_a^i$ a real vector
\cite{Fer,Gio}.

The non-vanishing Poisson brackets between the gravitational
variables are \cite{Kras,Martin}:
\begin{equation}\label{Poisson}
\{A_a^i(x),\tilde{E}^b_j(y)\}=\beta
l^2\delta^i_j\delta^b_a\delta^{(3)}(x-y),\end{equation} where
$\delta^{(3)}(x)$ is the Dirac delta on the spatial
three-manifold. We have adopted the convention that the
coordinates are dimensionless parameters \cite{LG}. The densitized
triad must then have the dimensions of the line element, i.e.
length squared. In addition, since $A_a^i$ is a connection, its
dimensionality must coincide with that of the derivative operator
and hence vanish.

Obviously (taking units in which $l=1$), a canonical set of
variables is given by $A_a^i$ and the scaled densitized triad
\begin{equation}\label{E}
E^a_i=\frac{\tilde{E}^a_i}{\beta}.\end{equation} The system has
three types of constraints: the Gauss, vector (or diffeomorphism)
and Hamiltonian (or scalar) ones \cite{book}. The kinematical
constraints adopt exactly the same expression in terms of any of
the pairs $(A_a^i,E^a_i)$:
\begin{eqnarray}\label{Gauss}
{\cal G}_i& \equiv & {\cal D}_aE^a_i=\partial_a
E^a_i+\epsilon_{ij}^{\;\;\;k}A_a^jE^a_k=0,\\ \label{diffeo} {\cal
V}_a&\equiv & E^b_i {\cal F}_{ba}^i =0.\end{eqnarray} Here, ${\cal
F}_{ab}^i=2\partial_{[a}A_{b]}^i+\epsilon^i_{\;jk}A_{a}^jA_{b}^k$
is the curvature of the connection $A_{a}^i$. As a result of the
invariance of the kinematical structure under changes of the value
of $\beta$, the predictions of quantum geometry (like, e.g., the
spectrum of the area operator, which can be constructed
exclusively from the densitized triad $\tilde{E}^a_i$) turn out to
depend on the Immirzi parameter \cite{Gio}. The quantum physics
displays therefore an ambiguity.

In order to show that the freedom in the choice of the Immirzi
parameter cannot be extended to a function on the spatial
manifold, let us first consider the possibility that, in
expression (\ref{AB}), $\beta$ becomes a function of the spatial
point, $\beta=\beta(x)$. Since $(\tilde{E}^a_i,K_a^i)$ is a
canonical set of variables, it is obvious that the same applies to
$(E^a_i,\beta K_a^i)$. The problem, however, is that the
transformation from the latter of these sets to the pair
$(A_a^i,E^a_i)$ is not canonical anymore when $\beta$ is not
constant.

Let us prove this statement. Since $A_a^i+\beta
K_a^i=\Gamma_a^i(\tilde{E})$ is just a function of the densitized
triad, the above transformation is canonical if and only if the
connection $\Gamma_a^i(\tilde{E})$ is the gradient of a generating
function $G$ of $E^a_i$, namely $\Gamma_a^i(\tilde{E})=\delta
G/\delta E^a_i$ \cite{book}. This would be the case if
$\Gamma_a^i(\tilde{E})$ coincided with the connection
$\Gamma_a^i(E)$ compatible with the scaled triad $E^a_i$, because
then $G=\int d^3x \Gamma_a^i(E) E^a_i$ (see page 81 of
\cite{book}). In fact, it is not difficult to check that a
variation $\delta E^a_i$ of the analyzed triad leads to a change
in its connection such that
\begin{equation} \label{var} 2E^a_i
\delta\Gamma_a^i(E)=\partial_a\left[\eta^{abc}
h_{cd}(E)E_b^i\delta E^d_i\right],\end{equation} where
$\eta^{abc}$ is the Levi-Civit\`{a} tensor density on the spatial
manifold and $h_{ab}(E)$ is the three-metric constructed with the
densitized triad $E^a_i$, i.e. $h_{ab}(\tilde{E})=\beta
h_{ab}(E)$. Since $E^a_i \delta\Gamma_a^i(E)$ is a total
derivative, it follows that $G$ is an acceptable generating
function.

However, the connections $\Gamma_a^i(\tilde{E})$ and
$\Gamma_a^i(E)$ differ when $\beta$ has a spatial dependence. In
other words, $\Gamma_a^i$ is invariant only under constant scale
transformations of the triad. From expression (\ref{spin}), it is
easy to obtain that
\begin{equation}\label{Gamma}
\Gamma_a^i(E)=\Gamma_a^i(\tilde{E})-\frac{1}{2}\epsilon_{\;\;\;k}^{ij}
E_a^kE^b_{j}\partial_b\ln{\beta}.\end{equation} The last factor is
not the gradient of a function of $E^a_i$ and, as a consequence,
neither is $\Gamma_a^i(\tilde{E})$. This can also be seen by
realizing that, from equation (\ref{var}) evaluated at
$\tilde{E}^a_i$,
\begin{equation} E^a_i
\delta\Gamma_a^i(\tilde{E})=\frac{\tilde{E}^a_i\delta\Gamma_a^i
(\tilde{E})}{\beta}= \frac{1}{2\beta}\partial_a\left[\beta
\eta^{abc} h_{cd}(E)E_b^i\delta E^d_i\right],\end{equation} which
is not a total derivative unless $\beta$ is constant.

In conclusion, when $\beta$ is spatially dependent, the set
$(A_a^i,E^a_i)$ is not canonical. Then, since the components of
the densitized triad commute between themselves and, therefore,
also with $\Gamma_a^i(\tilde{E})$, the Poisson brackets of the
components of $A_a^i$ cannot identically vanish.

Clearly, the only way to arrive at a SO(3) connection canonically
conjugate to $E^a_i$ is to replace $A_a^i$ with
\begin{equation}\label{Anu}
\overline{A}_a^{\,i}=\Gamma_a^i(E)-\beta K_a^i.\end{equation} Let
us now analyze whether the canonical transformation from the
Ashtekar-Barbero variables $(\underline{A}_a^{\,i},\tilde{E}^a_i)$
to the set $(\overline{A}_a^i,E^a_i)$ preserves the kinematical
structure. We will denote by $\overline{\cal D}_a$ and
$\underline{\cal D}_a$ the derivative operators defined by the
connections $\overline{A}_a^{\,i}$ and $\underline{A}_a^i$, and
their respective curvatures by $\overline{\cal F}_{ab}^i$ and
$\underline{\cal F}_{ab}^i$. Besides, $\overline{D}_a$ and
$\underline{D}_a$ will denote the derivative operators compatible
with the triads $E^a_i$ and $\tilde{E}^a_i$. For instance,
\begin{equation}\label{com}
\overline{D}_aE^b_i=\overline{\nabla}_aE^b_i+\epsilon_{ij}
^{\;\;\;k}\Gamma_a^j(E)E^a_k=0,\end{equation} where
$\overline{\nabla}_a$ is the covariant derivative of the metric
$h_{ab}(E)$, with no action on internal indices. We then have
\begin{equation}
{\cal G}_i\equiv \underline{\cal D}_a \tilde{E}^a_i\doteq
-\epsilon_{ij}^{\;\;\;k}K_a^j\tilde{E}^a_k=-\beta\epsilon_{ij}
^{\;\;\;k}K_a^jE^a_k\doteq \overline{\cal D}_a
E^a_i.\end{equation} The symbol $\doteq$ stands for equalities
where we have employed the compatibility equation (\ref{com}) or
its analogue $\underline{D}_a\tilde{E}^b_i=0$. So, the Gauss
constraint remains invariant under our canonical transformation:
it has the same expression for every possible function $\beta(x)$,
including the case $\beta=1$. In fact, this is consistent with the
expectation that $\overline{A}_a^{\,i}$ and $E^a_i$ should vary as
a connection and an internal vector, respectively, under SO(3)
transformations.

We now consider the effect of our change of variables in the
diffeomorphism constraint. It is known that this constraint can be
written as \cite{book,ijmp}:
\begin{equation}
{\cal V}_a\equiv\, \tilde{E}^b_i\underline{\cal F}_{ba}^i\doteq
{\cal G}_i
K^i_a+\underline{D}_b\left(\tilde{E}^c_iK^i_c\delta^b_a-
\tilde{E}^b_iK^i_a\right).\end{equation} In the above equality, in
addition to the compatibility of the connection $\Gamma_a^i$ with
the triad, we have taken into account the Bianchi identities,
which imply $\tilde{E}^b_i\underline{F}_{ba}^i=0$. Here,
$\underline{F}_{ab}^i$ is the curvature of
$\Gamma_a^i(\tilde{E})$. Using the analogue identity $E^b_i
\overline{F}_{ba}^i=0$ for the curvature of $\Gamma_a^i(E)$ and
the compatibility equation (\ref{com}), one similarly obtains
\begin{equation} \overline{\cal V}_a\equiv E^b_i
\overline{\cal F}_{ba}^i \doteq\beta {\cal
G}_iK^i_a+\overline{D}_b\left(\beta E^c_iK^i_c\delta^b_a- \beta
E^b_iK^i_a\right).\end{equation}

Note that the expression of the diffeomorphism constraint would
remain invariant only if ${\cal V}_a$ coincided with
$\overline{\cal V}_a$. The discrepancy between the action of the
derivative operators $\overline{D}_a$ and $\underline{D}_a$ comes
from the difference between the Christoffel symbols
$\Gamma^b_{ac}$ associated with the three-metrics $h_{ab}$. Using
the definition of these symbols \cite{ijmp,Wald} and
$h_{ab}(\tilde{E})=\beta h_{ab}(E)$, it is a simple exercise to
check that
\begin{equation}
\Gamma^b_{ac}(E)=\Gamma^b_{ac}(\tilde{E})+\frac{1}{2}\left[
h_{ac}(\tilde{E})h^{bd}(\tilde{E})-\delta^b_a\delta^d_c-
\delta^d_a\delta^b_c\right]\partial_d\ln{\beta}.\end{equation} A
straightforward calculation leads then to the result
\begin{equation}\label{rel} {\cal V}_a \doteq \overline{\cal V}_a
+{\cal G}_i\left[(1-\beta)K^i_a-\frac{1}{2}\epsilon_{\;\;\;k}
^{ij} E_{_{_{\!\!\!\!\!\!\sim}}\;a}^k
\tilde{E}_j^b\partial_b\ln{\beta}\right]-\tilde{E}^b_iK_b^i
\partial_a\ln{\beta},\end{equation} where we have employed
\begin{equation} {\cal G}_i \epsilon_{\;\;\;k}^{ij}
E_{_{_{\!\!\!\!\!\!\sim}}\;a}^k
\tilde{E}_j^b\doteq\tilde{E}^c_iK^i_d\left[ h_{ac}(\tilde{E})
h^{bd}(\tilde{E})-\delta^d_a\delta^b_c\right].\end{equation}
Therefore, when $\beta$ has a spatial dependence, ${\cal V}_a$ and
$\overline{\cal V}_a$ differ even modulo the Gauss constraint.

The reason of this discrepancy can be traced back to the fact that
the conformal transformation that relates the frames defined by
the densitized triads $\tilde{E}^a_i$ and $E^a_i$ is inhomogeneous
unless $\beta$ is constant. The vector constraint ${\cal V}_a$
generates diffeomorphisms on the spatial sections with line
element $h_{ab}(\tilde{E})dx^adx^b$, i.e. in the frame defined by
$\tilde{E}^a_i$. This frame is related to that associated with
$E^a_i$ by the (squared) conformal factor $\beta$. When $\beta$ is
spatially dependent, it is affected by changes of coordinates.
This is reflected by the last factor appearing in equation
(\ref{rel}). As a result of this term, the generators of
diffeomorphisms in the two analyzed frames are different [even
modulo SO(3) gauge]. It is worth noting that the extra term is
proportional to the densitized trace of the extrinsic curvature,
$\tilde{E}^a_iK_a^i$. Such a trace generates scale transformations
of the densitized triad accompanied by an inverse scaling of
$K_a^i$ \cite{TT}.

Accepting that the physically relevant frame is that of the
Ashtekar-Barbero formulation, we have thus seen that the
kinematical structure does not remain invariant under the analyzed
canonical transformations unless the Immirzi parameter is
constant. This precludes the possibility of generalizing the
Immirzi ambiguity to the extent of a spatial function, instead of
a real number.

That the extension of the Immirzi ambiguity finds an obstruction
in the vector constraint may be realized by considering the area
operator. If we insist that all geometrical quantities be measured
in the Ashtekar-Barbero frame, the area of a spatial surface $S$
(with unit normal $n_a$) in terms of the triad $E^a_i$ will be
given by $\int_S d^2x\beta(x)\sqrt{n_a n_b E^a_iE^{bi}}$. Since
the change from the Ashtekar-Barbero to the set of variables
$(\overline{A}_a^i,E^a_i)$ is canonical and preserves the
expression of the Gauss constraint, one can parallel the
construction of the area operator presented in reference
\cite{AL}, but adopting an $\overline{A}_a^i$-connection
representation (or its corresponding loop representation) instead
of the conventional representation of the Ashtekar-Barbero
formulation. The result would be an operator $\overline{\sc A}_S$
of the following form, compared with the standard area operator
$\underline{\sc A}_S$ of the $\underline{A}_a^i$ representation:
\begin{equation}\label{area}
\overline{\sc A}_S=\frac{l^2}{2}\sum_{v\in
S}\beta(v)\sqrt{-\overline{\Delta}_{S,v}},\hspace*{1.cm}
\underline{\sc A}_S=\frac{l^2}{2}\sum_{v\in
S}\sqrt{-\underline{\Delta}_{S,v}}.\end{equation} Here,
$\overline{\Delta}_{S,v}$ are the direct counterpart of the vertex
operators $\underline{\Delta}_{S,v}$ discussed by Ashtekar and
Lewandowski \cite{AL}, but now defined in the $\overline{A}_a^i$
representation. The sum, which in principle is over all the points
$v$ of $S$, becomes finite when the operator acts on cylindrical
functions, because the non-vanishing contributions come from the
isolated intersections of $S$ with the vertices of the graph
associated with the cylindrical state \cite{AL}.

However, note that, on cylindrical functions, the operator
$\overline{\sc A}_S$ depends on the spatial position of the
vertices $v$ through the function $\beta$. This dependence is
incompatible with diffeomorphism invariance.

An alternative manner to prove the incompatibility of
$\overline{\sc A}_S$ with the vector constraint is to show that
they do not commute. It is clear that $\overline{\sc A}_S$
commutes at least with the Gauss constraint, because the vertex
operators are gauge invariant. In the
$\underline{A}_a^i$-connection representation, on the other hand,
the vertex operators must commute with the quantum version of the
constraint ${\cal V}_a$ inasmuch as $\underline{\sc A}_S$ has a
well-defined action on the space of diffeomorphism-invariant
states. Choosing the same operator representation for
$\overline{\cal V}_a$ in the $\overline{A}_a^i$ formalism as for
${\cal V}_a$ in the Ashtekar-Barbero representation, and employing
that the area operator is a linear combination of vertex
operators, it follows that the commutator of $\overline{\sc A}_S$
and $\overline{\cal V}_a$ vanishes. Taking then into account
expression (\ref{rel}), we conclude that, when $\beta$ is
spatially dependent, $\overline{\sc A}_S$ would have to commute
with $E^a_iK_a^i$ in order to be a meaningful operator on
diffeomorphism-invariant states. Nonetheless, the considered
commutator must in fact differ from zero on the grounds that the
area is a functional of $E^a_i$ that scales like this triad and
that the densitized trace of the extrinsic curvature actually
generates scale transformations of $E^a_i$.

Summarizing, we have shown that the Immirzi ambiguity, that arises
in the loop quantization of general relativity, cannot be
generalized from a freedom in a constant parameter to the extent
of a function that depends on the spatial position. If we simply
allow the coefficient of the extrinsic curvature in the
Ashtekar-Barbero connection to be a spatially dependent function,
no scaling of the triad can be found that provides a canonically
conjugated variable. To attain a transformation that preserves the
canonical Poisson brackets one must, in addition, change the SO(3)
connection $\Gamma_a^i$ contained in the Ashtekar-Barbero
connection so that it becomes compatible with the scaled
densitized triad. The new set of canonical gravitational variables
leads to a connection formalism in which the expression of the
Gauss constraint remains unaffected, but the form of the vector
constraint is altered. This change in the kinematical structure of
the formulation of general relativity in terms of connections
spoils the compatibility of the geometrical operators, such as the
area operator, with the diffeomorphism invariance of the theory
and, in general, precludes the appearance of an extended Immirzi
ambiguity reflecting a spatial dependence.

The author is greatly thankful to J F Barbero G for enlightening
conversations and to L J Garay and T Thiemann for helpful comments
and discussions. This work was supported by the Spanish MCYT under
the research project no. BFM2001-0213.

\end{document}